\documentclass[10pt]{article}

\usepackage{graphicx}

\begin{document}

\title{LRS Bianchi Type I Models with Anisotropic Dark Energy and Constant Deceleration Parameter}
\date{} 
\author{\"{O}zg\"{u}r Akarsu\footnote{\"{O}zg\"{u}r Akarsu was supported in part by The Scientific and Technological Research Council of Turkey (T\"{U}B{\.I}TAK).} \and Can Battal K{\i}l{\i}n\c{c}}
\maketitle
\begin{center}
\vskip-1cm
\textit{Ege University, Faculty of Science, Dept. of Astronomy and Space Sciences, 35100 Bornova, {\.I}zmir/Turkey.} \textbf{E-mail:} \textit{ozgur.akarsu@mail.ege.edu.tr} (\textit{oa230@cam.ac.uk}), \textit{can.kilinc@ege.edu.tr}
\end{center}

\begin{abstract}
Locally rotationally symmetric (LRS) Bianchi Type I cosmological models are examined in the presence of dynamically anisotropic dark energy and perfect fluid. We assume that the dark energy (DE) is minimally interacting, has dynamical energy density, anisotropic equation of state parameter (EoS). The conservation of the energy-momentum tensor of the DE is assumed to consist of two separately additive conserved parts. A special law is assumed for the deviation from isotropic EoS, which is consistent with the assumption on the conservation of the energy-momentum tensor of the DE. Exact solutions of Einstein's field equations are obtained by assuming a special law of variation for the mean Hubble parameter, which yields a constant value of the deceleration parameter. Geometrical and kinematic properties of the models and the behaviour of the anisotropy of the dark energy has been carried out. The models give dynamically anisotropic expansion history for the universe that allows to fine tune the isotropization of the Bianchi metric, hence the CMB anisotropy.

%\keywords{anisotropic dark energy \and dynamical dark energy \and LRS Bianchi Type I \and constant deceleration parameter}
\end{abstract}
\section{Introduction}
\label{intro}
One of the most successful attempt to resolve the problems of standard Big Bang cosmology such as homogeneity, isotropy, and flatness of the universe is the inflationary paradigm, characterised by an epoch of accelerated expansion, "inflation", in the very early universe \cite{{Guth81},{Sato},{Linde},{Albrecht}}. During the inflationary epoch, quantum fluctuations are highly amplified, their wavelengths are stretched to outside the Hubble horizon and inevitably, superhorizon fluctuations are generated. These fluctuations become classical after crossing the event horizon and are coherent on what appear to be superhorizon scales at decoupling \cite{{Parker},{Birrell},{Mukhanov81},{Hawking},{Guth82},{Starobinsky},{Bardeen},{Mukhanov92},{Peiris},{Liddle}}. After the end of inflation, they re-enter the horizon, and seed the matter and the radiation fluctuations observed in the universe. These primordial fluctuations are Gaussian, adiabatic and nearly scale-invariant in the majority of inflation models and even a slight deviation from these properties can strongly constrain the assumptions in cosmological models \cite{{Peiris},{Liddle}}. Since the linearity of the cosmic microwave background (CMB) anisotropy preserves the basic properties of these primordial fluctuations, the CMB radiation anisotropy is a promising tool for testing these properties \cite{{Peiris}}. However, it is known that the observed quadrupole ($l=2$, which is the largest scale that can be observed) amplitude has a lower value than the quadrupole expected from a best-fit $\rm{\Lambda}$-dominated cold dark matter ($\rm{\Lambda}$CDM) standard model to the entire power spectrum since the first data of the differential microwave radiometer (COBE/DMR) appeared in 1992 \cite{{Smoot},{Bennett}}. This anomaly was confirmed with the high resolution data provided by the first year (2003) and 3-year (2006) Wilkinson Microwave Anisotropy Probe (WMAP, WMAP3) \cite{{Hinshaw03},{Hinshaw07}}. Most recently, the 5-year WMAP data did not improve on the quadrupole, it still seems to be outside of the fit \cite{{Nolta},{Hinshaw08}}. Today, this low value of the quadrupole seems inescapable and the main issue now is to elucidate possible explanations \cite{{Eriksen}}. The proposed four possible candidates are that this anomaly is due to a systematic error, a pure statistical fluke, astrophysical (i.e. unexpected foreground) or cosmological reasons (e.g. nontrivial spatial geometry of the universe) \cite{{Eriksen},{Copi}}. Although the correct explanation is still unknown, the first three candidates seem unlikely \cite{{Hinshaw08},{Eriksen},{Copi},{Spergel},{Gold},{Hill}}. If it is indeed due to cosmological reasons, since the lowest multipoles represent especially the scale of the horizon at approximately dark energy (DE) domination, it seems natural to associate the low value of the quadrupole with the nature of the DE.

It is known that Bianchi universe anisotropies give rise to CMB anisotropies depending on the model type \cite{{Ellis}}. Recently, Campanelli et al. showed that allowing large-scale spatial geometry of the universe to be plane-symmetric with eccentricity (regardless of the origin) at decoupling of order $10^{-2}$ can bring the quadrupole amplitude in accordance with observations without affecting higher multipoles of the angular power spectrum of the temperature anisotropy \cite{{Campanelli06},{Campanelli07}}. Although they are controversial, it should be mentioned here that there are also some independent indications of a symmetry axis in the large-scale geometry of the universe, coming from the analysis of spiral galaxies in the Sloan Digital Sky Survey (SDSS) \cite{{Longo}} and from the analysis of polarization of electromagnetic radiation propagating over cosmological distances \cite{{Nodland},{Hutsemekers},{Battye},{Borguet08a},{Borguet08b},{Joshi},{Rakic07},{Rakic06}}. A spatially ellipsoidal geometry of the universe can be described with Bianchi type metrics. However, Bianchi type I, V, VII models isotropize at late times even for ordinary matter, and the possible anisotropy of the Bianchi metrics necessarily die away during the inflationary era \cite{{Ellis},{Bonometto}}. In fact this isotropization of the Bianchi metrics is due to the implicit assumption that the DE is isotropic in nature. If the implicit assumption that the pressure of the DE is direction independent is relaxed, the isotropization of the Bianchi metrics can be fine tuned to generate arbitrary ellipsoidality (eccentricity). Therefore, the CMB anisotropy can also be fine tuned, since the Bianchi universe anisotropies determine the CMB anisotropies. The price of this property of DE is a violation of the null energy condition (NEC) since the DE crosses the Phantom Divide Line (PDL), in particular depending on the direction \cite{{Rodrigues}}.

However, it has been known since the 1980s that such energy components might occur and their role as possible DE candidates was raised by Caldwell at the end 1990s (see \cite{{Carroll}} for further references). In theory, despite the observational constraints, extensions of general relativity are the prime candidate class of theories consistent with PDL crossing \cite{{Nesseris}}. On the other hand, while the current cosmological data from SNIa (Supernova Legacy Survey, Gold sample of Hubble Space Telescope) \cite{{Riess},{Astier}}, CMB (WMAP, BOOMERANG) \cite{{Komatsu},{MacTavish}} and large scale structure (SDSS) \cite{{Eisenstein}} data rule out the $w\ll-1$, they mildly favor dynamically evolving DE crossing the PDL (see \cite{{Carroll},{Nesseris}}, \cite{{Zhao},{Copeland}} for theoretical and observational status of crossing the PDL).

Recently, Rodrigues \cite{{Rodrigues}} and Koivisto \& Mota \cite{{Koivisto08a},{Koivisto08b}} have investigated cosmological models with anisotropic equation of state (EoS). Rodrigues has constructed a Bianchi type-I $\rm{\Lambda}$CDM cosmological model with a DE component which is non-dynamical but wields anisotropic vacuum pressure in two ways: i) by implementing of anisotropic vacuum pressure consistent with energy-momentum tensor conservation; ii) by implementing a Poisson structure deformation between canonical momenta such that rescaling of the scale factor is not violated \cite{{Rodrigues}}. He suggests to fine tune the DE so as to not wipe out the anisotropic imprints in the inflationary epoch. On the other hand, Koivisto \& Mota have proposed a different approach to resolve CMB anisotropy problem; even if the CMB formed isotropically at early time, it could be distorted by the direction dependent acceleration of the later universe in such a way that it appears to us anomalous at the largest scales. They have investigated a cosmological model containing a DE component which has a non-dynamical anisotropic EoS and interacts with the perfect fluid component. They have also suggested that cosmological models with anisotropic EoS can explain the quadrupole problem and can be tested by SNIa data \cite{{Koivisto08a},{Koivisto08b}}.

In reference \cite{{Mota}} Mota et al. have concluded that even though a perfect fluid representation might ultimately turn out to be a phenomenologically sufficient description of all the observational consequences of DE, imperfectness in DE cannot be excluded \cite{{Mota}}. Although there is compelling evidence that the expansion of universe is speeding up, we are far from understanding of the nature of the DE which is thought be the reason for this behaviour \cite{{Carroll},{Burdyuzha},{Turner}}. Hence, we should examine models with anisotropic dark energy, in order to determine what possible new physical consequences they might give rise to, and if for no other reason than to rule such models out.

The above discussions lead us to examine the physical behaviours of the two-fluid, particularly $\rm{\Lambda}$CDM, cosmological models with an anisotropic DE component. Models of DE are conveniently characterized by the EoS parameter $w=p/\rho$ which is not necessarily constant, where $\rho$ is the energy density and $p$ is the pressure \cite{{Carroll}}. However, while energy density is a scalar quantity, pressure is a vectorial quantity, and consequently the EoS parameter of DE may be determined separately on each spatial axis in a consistent way with the conservation of energy-momentum tensor. Hence, we consider a phenomenological parametrization of minimally interacting DE in terms of its time-dependent deviation-free equation of state parameter $w^{\mathrm{\mathrm{(de)}}}(t)$ and deviation parameters $[\delta(t),\gamma(t),\gamma(t)]$. Since such a parametrization yields an anisotropic expansion which is not compatible with the Robertson-Walker (RW) metric, in section 2 we have used the locally rotationally symmetric Bianchi Type I (LRS Bianchi-I) metric which generalizes the flat RW metric in an axially symmetric way and is compatible with our parametrization. The other component has assumed to be a perfect fluid (dark matter or ordinary matter). With our assumptions, the conservation of the energy-momentum tensor implies dynamically anisotropic DE except for very special solutions. We have obtained exact solutions for the equations by assuming a special dynamic for the anisotropy of the dark the energy and a special law of variation for the mean Hubble parameter in Bianchi metrics, which yields a constant value of deceleration parameter \cite{{Singh06},{Kumar},{Singh07}}. The assumption on the mean Hubble parameter allows us to determine the scale factors exactly, as well as to examine the behaviour of the anisotropy of DE and other cosmological parameters of such a universe.

\section{Model and field equations}
\label{sec:1}
The spatially homogenous, anisotropic and LRS Bianchi-I space-time is described by the line element
\begin{equation}
ds^{2}=dt^{2}-A(t)^{2}dx^{2}-B(t)^{2}(dy^{2}+dz^{2})
\end{equation}
where $A(t)$ and $B(t)$ are the scale factors (metric tensors) and functions of the cosmic time $t$. In natural units ($8\pi G=1$ and $c=1$), the field equations, in the case of a mixture of the perfect fluid and the anisotropic DE components, are
\begin{equation}
G_{\mu\nu}=R_{\mu\nu}-\frac{1}{2}Rg_{\mu\nu}=-{T^{\mathrm{(m)}}}_{\mu\nu}-{T^{\mathrm{(de)}}}_{\mu\nu}
\end{equation}
with
\begin{eqnarray}
{{T^{\mathrm{(m)}}}_{\nu}}^{\mu}=\mathrm{diag}[\rho^{\mathrm{(m)}},-p^{\mathrm{(m)}},-p^{\mathrm{(m)}},-p^{\mathrm{(m)}}]\\
\nonumber =\mathrm{diag}[1,-w^{\mathrm{(m)}},-w^{\mathrm{(m)}},-w^{\mathrm{(m)}}]\rho^{\mathrm{(m)}}
\end{eqnarray}
and
\begin{eqnarray}
{{T^{\mathrm{(de)}}}_{\nu}}^{\mu}=\mathrm{diag}[\rho^{\mathrm{(de)}},-{p_{x}}^{\mathrm{(de)}},-{p_{y}}^{\mathrm{(de)}},-{p_{z}}^{\mathrm{(de)}}]\\
\nonumber =\mathrm{diag}[1,-{w_{x}}^{\mathrm{(de)}},-{w_{y}}^{\mathrm{(de)}},-{w_{z}}^{\mathrm{(de)}}]\rho^{\mathrm{(de)}}\\
\nonumber =\mathrm{diag}[1,-(w^{\mathrm{(de)}}+\delta),-(w^{\mathrm{(de)}}+\gamma),-(w^{\mathrm{(de)}}+\gamma)]\rho^{\mathrm{(de)}}
\end{eqnarray}
where $g_{\mu\nu}u^{\mu}u^{\nu}=1$; $u^{\mu}=(1,0,0,0)$ is the four-velocity vector; $R_{\mu\nu}$ is the Ricci tensor; $R$ is the Ricci scalar; $\rho^{\mathrm{(m)}}$ and $\rho^{\mathrm{(de)}}$ are the energy densities of the perfect fluid and DE components respectively; $w^{\mathrm{(m)}}$ is the EoS parameter of the perfect fluid and we will call $w^{\mathrm{(de)}}$ the deviation-free EoS parameter of the DE. Here, the perfect fluid represents the ordinary matter or cold dark matter, thus $w^{(m)}\geq0$. ${w_{x}}^{\mathrm{(de)}}$, ${w_{y}}^{\mathrm{(de)}}$ and ${w_{z}}^{\mathrm{(de)}}$ are the directional EoS parameters of the DE on $x$, $y$ and $z$ axes respectively. ${w_{y}}^{\mathrm{(de)}}$ and ${w_{z}}^{\mathrm{(de)}}$ are set to be equal which is convenient with the metric given in (1). $\delta$ and $\gamma$ are deviations from the deviation-free EoS parameter (hence the deviation-free pressure) of the DE respectively on $x$ axis and $y$ and $z$ axes. 
In a comoving coordinate system, Einstein's field equations (2), for the anisotropic LRS Bianchi-I space-time (1), in case of (3) and (4), read as 
\begin{equation}
\frac{\dot{B}^{2}}{B^{2}}+2\frac{\dot{A}}{A}\frac{\dot{B}}{B}=\rho^{\mathrm{(m)}}+\rho^{\mathrm{(de)}},
\end{equation}
\begin{equation}
\frac{\dot{B}^{2}}{B^{2}}+2\frac{\ddot{B}}{B}=-w^{\mathrm{(m)}}\rho^{\mathrm{(m)}}-(w^{\mathrm{(de)}}+\delta)\rho^{\mathrm{(de)}},
\end{equation}
\begin{equation}
\frac{\ddot{B}}{B}+\frac{\dot{B}}{B}\frac{\dot{A}}{A}+\frac{\ddot{A}}{A}=-w^{\mathrm{(m)}}\rho^{\mathrm{(m)}}-(w^{\mathrm{(de)}}+\gamma)\rho^{\mathrm{(de)}},
\end{equation}
where the over dot denotes derivation with respect to the cosmic time $t$. We have the following equation from the Bianchi identity,
\begin{equation}
{{G}^{\mu\nu}}_{;\nu}={{T^{\mathrm{(m)}}}^{\mu\nu}}_{;\nu}+{{T^{\mathrm{(de)}}}^{\mu\nu}}_{;\nu}=0,
\end{equation}
which yields
\begin{eqnarray}
{\dot{\rho}}^{\mathrm{(m)}}+\left(1+w^{\mathrm{(m)}}\right)\rho^{\mathrm{(m)}}\left(\frac{\dot{A}}{A}+2\frac{\dot{B}}{B}\right)\\
\nonumber +{\dot{\rho}}^{\mathrm{(de)}}+\left(1+w^{\mathrm{(de)}}\right)\rho^{\mathrm{(de)}}\left(\frac{\dot{A}}{A}+2\frac{\dot{B}}{B}\right)+\rho^{\mathrm{(de)}}\left(\delta\frac{\dot{A}}{A}+2\gamma\frac{\dot{B}}{B}\right)=0.
\end{eqnarray}
This equation, which is linearly dependent to Einstein field equations, also represents the conservation of the total energy momentum tensor.
\section{Solution of the field equations}
\label{sec:3}
We have initially 8 variables ($A$, $B$, $\rho^{\mathrm{(m)}}$, $w^{\mathrm{(m)}}$, $\rho^{\mathrm{(de)}}$, $w^{\mathrm{(de)}}$, $\delta$,$\gamma$) and four equations, three of which are linearly independent, namely three Einstein field equations (5-7) and the Bianchi identity. The system is thus initially undetermined and we need additional constraints to close the system.
We have assumed that the DE is minimally interacting, ${{T^{\mathrm{(de)}}}^{\mu\nu}}_{;\nu}=0$; thus due to the Bianchi identity (8) the perfect fluid component is also minimally interacting, ${{T^{\mathrm{(m)}}}^{\mu\nu}}_{;\nu}=0$. Hence, the Bianchi identity has been split into two separately additive conserved components; namely, the conservation of the energy-momentum tensor of the DE
\begin{equation}
{{T^{\mathrm{(de)}}}^{\mu\nu}}_{;\nu}={\dot{\rho}}^{\mathrm{(de)}}+\left(1+w^{\mathrm{(de)}}\right)\rho^{\mathrm{(de)}}\left(\frac{\dot{A}}{A}+2\frac{\dot{B}}{B}\right)+\rho^{\mathrm{(de)}}\left(\delta\frac{\dot{A}}{A}+2\gamma\frac{\dot{B}}{B}\right)=0
\end{equation}
and the conservation of the energy-momentum tensor of the perfect fluid component
\begin{equation}
{{T^{\mathrm{(m)}}}^{\mu\nu}}_{;\nu}={\dot{\rho}}^{\mathrm{(m)}}+\left(1+w^{\mathrm{(m)}}\right)\rho^{\mathrm{(m)}}\left(\frac{\dot{A}}{A}+2\frac{\dot{B}}{B}\right)=0.
\end{equation}
Once more we may split the conservation of the energy-momentum tensor of the dark-enery into two parts:
\begin{equation}
{{T^{\mathrm{(de)}}}^{\mu\nu}}_{;\nu}={{T^{'\mathrm{(de)}}}^{\mu\nu}}_{;\nu}+{{\tau^{\mathrm{(de)}}}^{\mu\nu}}_{;\nu}=0,
\end{equation}
where ${{\tau^{\mathrm{(de)}}}^{\mu\nu}}_{;\nu}$ is the last term of the ${{T^{\mathrm{(de)}}}^{\mu\nu}}_{;\nu}$ in (10) and arises due to the deviations from $w^{\mathrm{(de)}}$, and ${{T^{'\mathrm{(de)}}}^{\mu\nu}}_{;\nu}$ is the the deviation-free part of the ${{T^{\mathrm{(de)}}}^{\mu\nu}}_{;\nu}$ in (10). Now, we will do the following strong assumption,
\begin{equation}
{{\tau^{\mathrm{(de)}}}^{\mu\nu}}_{;\nu}=\rho^{\mathrm{(de)}}\left(\delta\frac{\dot{A}}{A}+2\gamma\frac{\dot{B}}{B}\right)=0,
\end{equation}
which also results in the deviation-free part of the ${{T^{\mathrm{(de)}}}^{\mu\nu}}_{;\nu}$ to be null, that is,
\begin{equation}
{{T^{'\mathrm{(de)}}}^{\mu\nu}}_{;\nu}={\dot{\rho}}^{\mathrm{(de)}}+\left(1+w^{\mathrm{(de)}}\right)\rho^{\mathrm{(de)}}\left(\frac{\dot{A}}{A}+2\frac{\dot{B}}{B}\right)=0,
\end{equation}
which looks like the conservation of the energy momentum tensor of a minimally interacting perfect fluid. According to (13) and (14) the behaviour of $\rho^{\mathrm{(de)}}$ is controlled by the deviation-free part of the EoS parameter of the DE ($w^{\mathrm{(de)}}$), but the deviations will affect $\rho^{\mathrm{(de)}}$ indirectly, since, as can be seen later, they affect the value of $w^{\mathrm{(de)}}$. If the deviation parameters are assumed to be constants, to assure ${{\tau^{\mathrm{(de)}}}^{\mu\nu}}_{;\nu}=0$ either $\delta$ and $\gamma$ are trivially null or the ratio of the expansion rate on $x$ axis to the expansion rate on $y$ axis is equal to $-2\gamma/\delta$ which is a very special case. On the other hand, we may get more general solutions, if $\delta$ and $\gamma$ are allowed to be function of the cosmic time $t$ and we constrained $\delta$ and $\gamma$ by assuming a special dynamic which is consistent with (13). The dynamic of the deviation parameter on the $x$ axis, $\delta(t)$, is assumed to be
\begin{equation}
\delta(t)=n\frac{2}{3}\frac{\dot{B}}{B}\left(\frac{\dot{A}}{A}+2\frac{\dot{B}}{B}\right)\frac{1}{\rho^{\mathrm{(de)}}},
\end{equation}
and thus from (13) the deviation parameter on the $y$ and $z$ axes, $\gamma(t)$, is found as
\begin{equation}
\gamma(t)=-n\frac{1}{3}\frac{\dot{A}}{A}\left(\frac{\dot{A}}{A}+2\frac{\dot{B}}{B}\right)\frac{1}{\rho^{\mathrm{(de)}}}.
\end{equation}
In such an assumption $\delta(t)$ and $\gamma(t)$ are dimensionless parameters, and $n$ is a dimensionless constant that parametrizes the amplitude of the deviation from $w^{\mathrm{(de)}}$ and can be given real values. The measure of the anisotropy of the DE may be given by $(\delta(t)-\gamma(t))/w^{\mathrm{(de)}}$ and it is null, which implies the DE is isotropic, when $n=0$.

The EoS parameter of the perfect fluid has been assumed to be constant,
\begin{equation}
w^{\mathrm{(m)}}=\frac{p^{\mathrm{(m)}}}{\rho^{\mathrm{(m)}}}=const;
\end{equation} 
while $w^{\mathrm{(de)}}$ is allowed to be a function of the cosmic time, since the current cosmological data from SNIa, CMB and large scale structures mildly favor dynamically evolving DE crossing the PDL as mentioned in the Section 1. Hence, since $w^{\mathrm{(de)}}(t)$ hasn't been constrained, we still need one more constraint to close the system. We imposed a law of variation for the Hubble parameter. The law of variation for the Hubble parameter which was initially proposed by Berman for RW space-time and yields a constant value of deceleration parameter \cite{{Singh06},{Kumar},{Singh07},{Singh08},{Berman83},{Berman88}}. Such a law of variation for Hubble parameter given by Berman is not inconsistent with observations \cite{{Kumar},{Singh07}} and is also approximately valid for slowly time varying deceleration parameter \cite{Singh08}. Recently Singh and Kumar proposed a similar law of variation for the Hubble parameter in anisotropic space-time metrics that yields a constant value of the deceleration parameter, and generated solutions for Bianchi Type-I \cite{{Kumar}}, LRS Bianchi Type-II \cite{{Singh06},{Singh07}}, Bianchi Type-V \cite{Singh08} metrics in General Relativity. According to the proposed law, the variation of the mean Hubble parameter for the LRS Bianchi-I metric may be given by
\begin{equation}
H=k(AB^{2})^{-\frac{m}{3}},
\end{equation}
where $k>0$ and $m\geq0$ are constants. The spatial volume is given by
\begin{equation}
V=a^{3}=AB^{2}
\end{equation}
where $a$ is the mean scale factor. The mean Hubble parameter $H$ for LRS Bianchi-I metric may be given by
\begin{equation}
H=\frac{\dot{a}}{a}=\frac{1}{3}\frac{\dot{V}}{V}=\frac{1}{3}\left(\frac{\dot{A}}{A}+2\frac{\dot{B}}{B}\right).
\end{equation}
The directional Hubble parameters in the directions of $x$, $y$ and $z$ respectively may be defined as
\begin{equation}
H_{x}\equiv\frac{\dot{A}}{A}\qquad \textnormal{and}\qquad H_{y}=H_{z}\equiv\frac{\dot{B}}{B}.
\end{equation}
(Since $H_{y}=H_{z}$, in the following they are represented by $H_{y,z}$.) The volumetric deceleration parameter is
\begin{equation}
q=-\frac{a\ddot{a}}{\dot{a}^{2}}.
\end{equation}
On integration, after equating (18) and (20), we get
\begin{equation}
AB^{2}=c_{1}e^{3kt}\qquad\textnormal{for}\quad m=0
\end{equation}
and
\begin{equation}
AB^{2}=\left(mkt+c_{2}\right)^{\frac{3}{m}}\qquad\textnormal{for}\quad m\neq 0.
\end{equation}
Here $c_{1}$ and $c_{2}$ are positive constants of integration. Using (18) with (23) for $m=0$, and with (24) for $m\neq 0$ mean Hubble paremeters are
\begin{equation}
H=k\qquad\textnormal{for}\quad m=0
\end{equation}
and
\begin{equation}
H=k(mkt+c_{2})^{-1}\qquad\textnormal{for}\quad m\neq0.
\end{equation}
Using (23-24) and (19) in (22) we get constant values for the deceleration parameter for the mean scale factor as:
\begin{equation}
q=m-1\qquad \textnormal{for}\quad m\neq 0
\end{equation}
and
\begin{equation}
q=-1\qquad \textnormal{for}\quad m=0.
\end{equation}
The sign of $q$ indicates whether the model accelerate or not. The positive sign of $q$ (i.e. $m>1$) corresponds to decelerating models whereas the negative sign $-1\leq q<0$ for $0\leq m< 1$ indicates acceleration and $q=0$ for $m=1$ correponds to expansion with constant velocity.

Using the deviation parameters (15) and (16), and the mean Hubble parameter (20) in the subtraction (6) from (7) we may get
\begin{equation}
\frac{d}{dt} \left(\frac{\dot{A}}{A}-\frac{\dot{B}}{B} \right)+\left( \frac{\dot{A}}{A}-\frac{\dot{B}}{B} \right) 3H=3nH^{2}
\end{equation}
after little manipulation. On integration of (29) by considering (25) and (26) we obtain
\begin{equation}
\frac{\dot{A}}{A}-\frac{\dot{B}}{B}=\lambda e^{-3kt}+nk\qquad \textnormal{for}\quad m=0,
\end{equation}
\begin{equation}
\frac{\dot{A}}{A}-\frac{\dot{B}}{B}=\frac{\lambda}{\left(mkt+c_{2}\right)^{\frac{3}{m}}}+\frac{3nk}{(3-m)(mkt+c_{2})}\qquad \textnormal{for}\quad m\neq 0\textnormal{ and } 3
\end{equation}
and
\begin{equation}
\frac{\dot{A}}{A}-\frac{\dot{B}}{B}=\frac{\lambda}{3kt+c_{2}}+\frac{nk\ln(3kt+c_{2})}{3kt+c_{2}}\qquad \textnormal{for}\quad m=3,
\end{equation}
where $\lambda$ is the real constant of integration. One can observe that $\lambda$ and $n$ are two parameters that parametrize the difference between the directional Hubble parameters. Now, we can find $A(t)$ and $B(t)$ explicitly for all $m$ values by using (30-32).
\subsection{Model for $m=0\quad(q=-1)$}
Using (30) we may get the ratios of the scale factors $A(t)/B(t)$, and on manipulating the result by using (23) we get the following exact expressions for the scale factors:
\begin{equation}
A(t)={c_{1}}^{1/3}\kappa^{2/3}e^{kt-\frac{2}{9}\frac{\lambda}{k}e^{-3kt}+\frac{2}{3}nkt},
\end{equation}
\begin{equation}
B(t)={c_{1}}^{1/3}\kappa^{-1/3}e^{kt+\frac{1}{9}\frac{\lambda}{k}e^{-3kt}-\frac{1}{3}nkt},
\end{equation}
where $\kappa$ is the positive constant of integration. The spatial volume of the universe is found as
\begin{equation}
V=c_{1}e^{3kt}.
\end{equation}
The directional Hubble parameters as defined in (21) are found as
\begin{equation}
H_{x}=k+\frac{2}{3}{\lambda}e^{-3kt}+\frac{2}{3}kn,
\end{equation}
\begin{equation}
H_{y,z}=k-\frac{1}{3}{\lambda}e^{-3kt}-\frac{1}{3}kn.
\end{equation}
The anisotropy parameter of the expansion $\Delta$ is defined as
\begin{equation}
\Delta\equiv\frac{1}{3}\sum_{i=1}^{3}\left(\frac{H_{i}-H}{H}\right)^{2},
\end{equation}
where $H_{i}$ (i=1,2,3) represents the directional Hubble parameters in the directions of $x$, $y$ and $z$ respectively. By using equations (25),(36) and (37) in (38) we get
\begin{equation}
\Delta=\frac{2}{9}{\frac{\left(\lambda{e^{-3kt}+nk}\right)^{2}}{{k}^{2}}}
\end{equation}
for the anisotropy of the expansion. The expansion scalar, defined by $\Theta\equiv{u^{i}}_{;i}$, is found as
\begin{equation}
\Theta =3k=3H.
\end{equation}
The shear scalar, defined by $\sigma^{2}\equiv\frac{1}{2}\sigma_{ij}\sigma^{ij}$ where $\sigma_{ij}=u_{i;j}+u_{j;i}-\frac{2}{3}g_{ij}{u^{k}}_{;k}$ is the shear tensor, is found as
\begin{equation}
\sigma^{2}=\frac{1}{3}\left({\lambda}e^{-3kt}+nk\right)^{2}.
\end{equation}
Using the scale factors in (11), the energy density of the perfect fluid is found as
\begin{equation}
\rho^{\mathrm{(m)}}(t) ={\rho^{\mathrm{(m)}}_{0}}{e^{-3k \left( 1+w^{\mathrm{(m)}} \right) t}}.
\end{equation}
The energy density of the DE may be found from (5) by using the scale factors and the energy density of the perfect fluid (42), and may be written in terms of $\Delta(t)$ as
\begin{equation}
\rho^{\mathrm{(de)}}(t) =3k^{2}\left(1-\frac{1}{2}\Delta(t)\right)-\rho^{\mathrm{(m)}}(t).
\end{equation}
Using the scale factors and (43) in (14) we get
\begin{equation}
w^{\mathrm{(de)}}(t)={\frac{3w^{\mathrm{(m)}}\rho^{\mathrm{(m)}}(t)+{\lambda}^{2}{e^{-6kt}}+k^{2}\left( 9-{n}^{2}\right)}{3\rho^{\mathrm{(m)}}(t)+{\lambda}^{2}{e^{-6kt}}+2{\lambda}kn{e^{-3kt}}-k^{2}\left(9-{n}^{2}\right)}}
\end{equation}
for the deviation-free EoS parameter of the DE.
Using the scale factors and (43) in (15) and (16) we may get deviation parameters as following,
\begin{equation}
\delta(t)={\frac{2{\lambda}kn{e^{-3kt}}+2{k}^{2}n(n-3)}{3\rho^{\mathrm{(m)}}(t)+{\lambda}^{2}{e^{-6kt}}+2{\lambda}kn{e^{-3kt}}-k^{2}\left(9-{n}^{2}\right)}},
\end{equation}
\begin{equation}
\gamma(t)={\frac{2{\lambda}kn{e^{-3kt}}+2{k}^{2}n(n+3/2)}{3\rho^{\mathrm{(m)}}(t)+{\lambda}^{2}{e^{-6kt}}+2{\lambda}kn{e^{-3kt}}-k^{2}\left(9-{n}^{2}\right)}}.
\end{equation}
\subsection{Physical behaviour of the model for $m=0\quad(q=-1)$}
For this model $q=-1$ and $dH/dt=0$, which implies the greatest value of the Hubble parameter and the fastest rate expansion of the universe. Thus, this model may represent the inflationary era in the early universe and the very late times of the universe.

The spatial volume $V$ is finite at $t=0$, expands exponentially as $t$ increases and becomes infinitely large at $t =\infty$. The directional Hubble parameters $H_{x}$ and $H_{y,z}$ are finite at $t=0$ and $t = \infty$. They deviate from the mean Hubble parameter $H$ due to $\lambda$ and $n$. While $\lambda$ is supporting (opposing) the expansion on the $x$ axis, it opposes (supports) the expansion on $y$ and $z$ axes. However it loses its effect exponentially by the cosmic time $t$. The anisotropy of the DE has a similar effect on the directional Hubble parameters, but its effect persists. While $n$ is supporting (opposing) the expansion on the $x$ axis, it opposes (supports) the expansion on $y$ and $z$ axes. The anisotropy of the DE doesn't always act so as to increase the anisotropy of the expansion. In fact, when the signs of $\lambda$ and $n$ are opposite the overall effect is to lower the expansion anisotropy.

The expansion scalar is constant throughout the evolution of the universe. The shear scalar is also finite at $t=0$ and tends to $n^{2}k^{2}/3$ as $t$ increases. If $-3/2<n<3$ all the axes will expand to infinitly large values as $t\rightarrow\infty$. On the other hand, the space-time exhibits a pancake type singularity for $n<-3/2$ and a cigar type singularity for $n>3$ at $t=\infty$. If $n=-3/2$ while the $x$ axis converges to a constant, $y$ and $z$ axes expand to infinitly large values, and it is vice versa if $n=3$.
\begin{figure}[ht]
\begin{minipage}[b]{0.49\linewidth}
\centering
\includegraphics[width=1\textwidth]{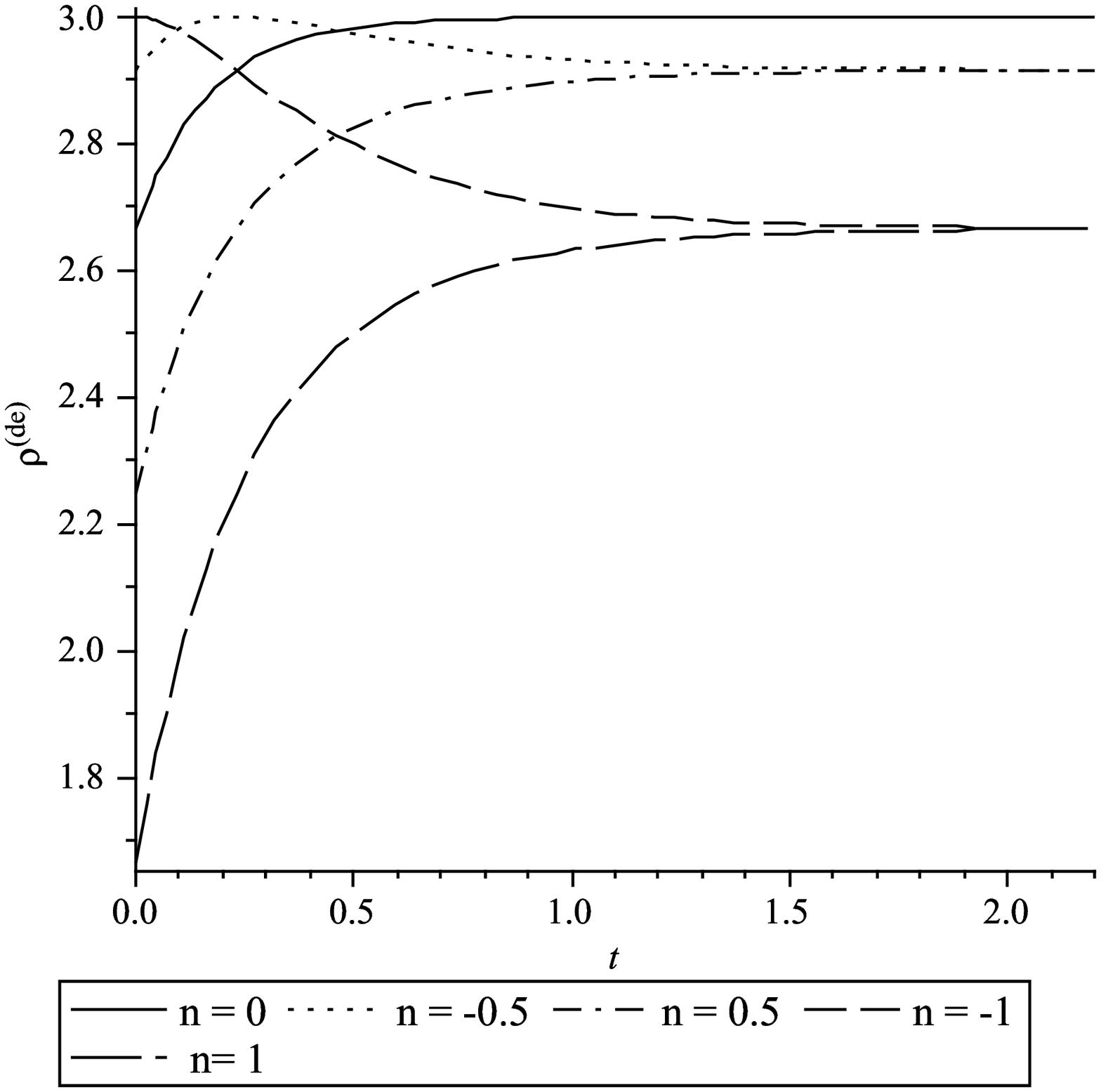}
\caption{$\rho^{\mathrm{(de)}}$ versus cosmic time $t$ for different values of $n$ in model $m=0$. $\lambda$ and $k$ have been chosen as $1$.}
\label{fig:rhom0}
\end{minipage}
\hspace{0.01\linewidth}
\begin{minipage}[b]{0.49\linewidth}
\centering
\includegraphics[width=1\textwidth]{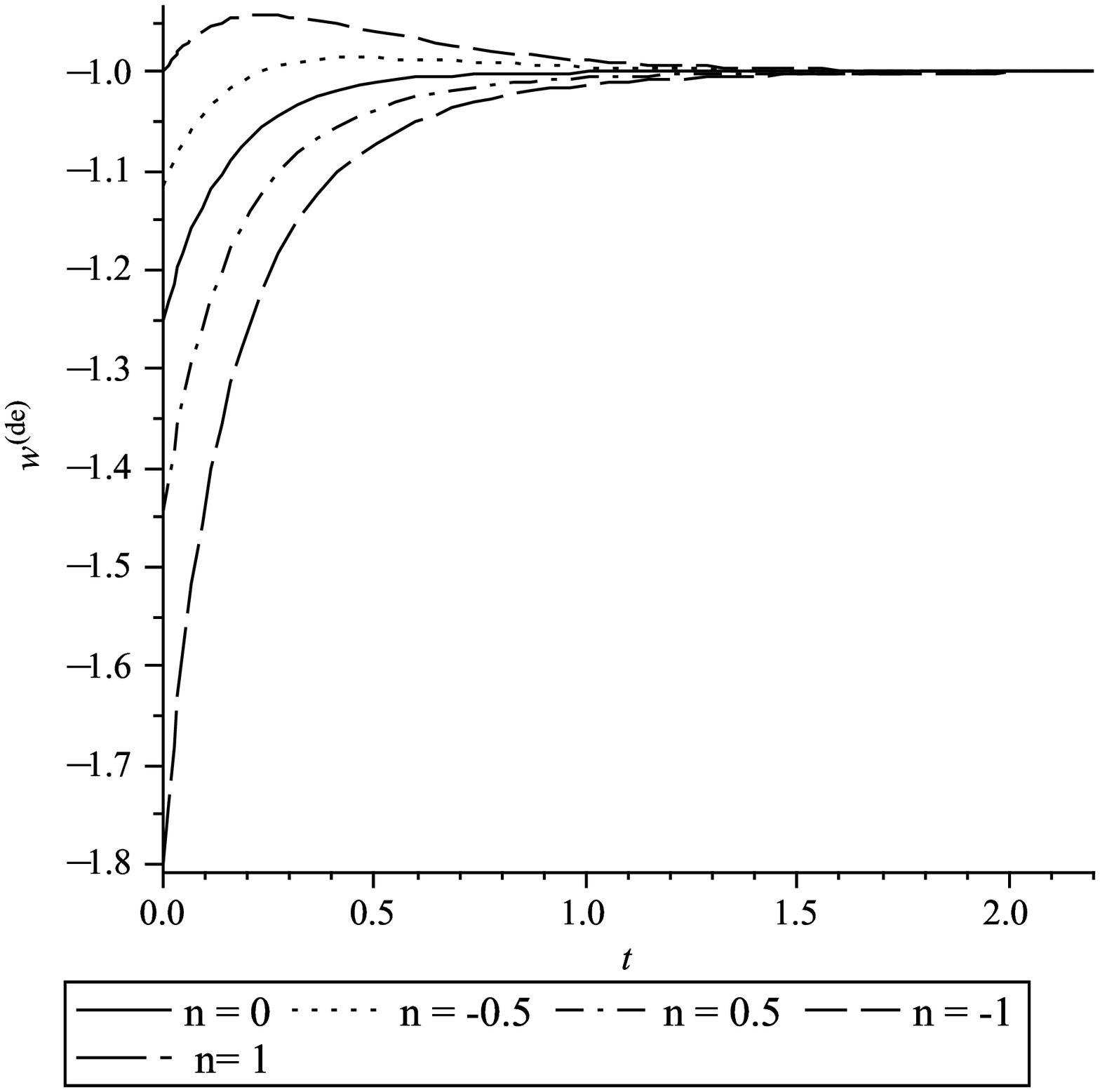}
\caption{$w^{\mathrm{(de)}}$ versus cosmic time $t$ for different values of $n$ in model $m=0$. $\lambda$ and $k$ have been chosen as $1$.}
\label{fig:wm0}
\end{minipage}
\end{figure}
\begin{figure}[ht]
\begin{minipage}[b]{0.49\linewidth}
\centering
\includegraphics[width=1\textwidth]{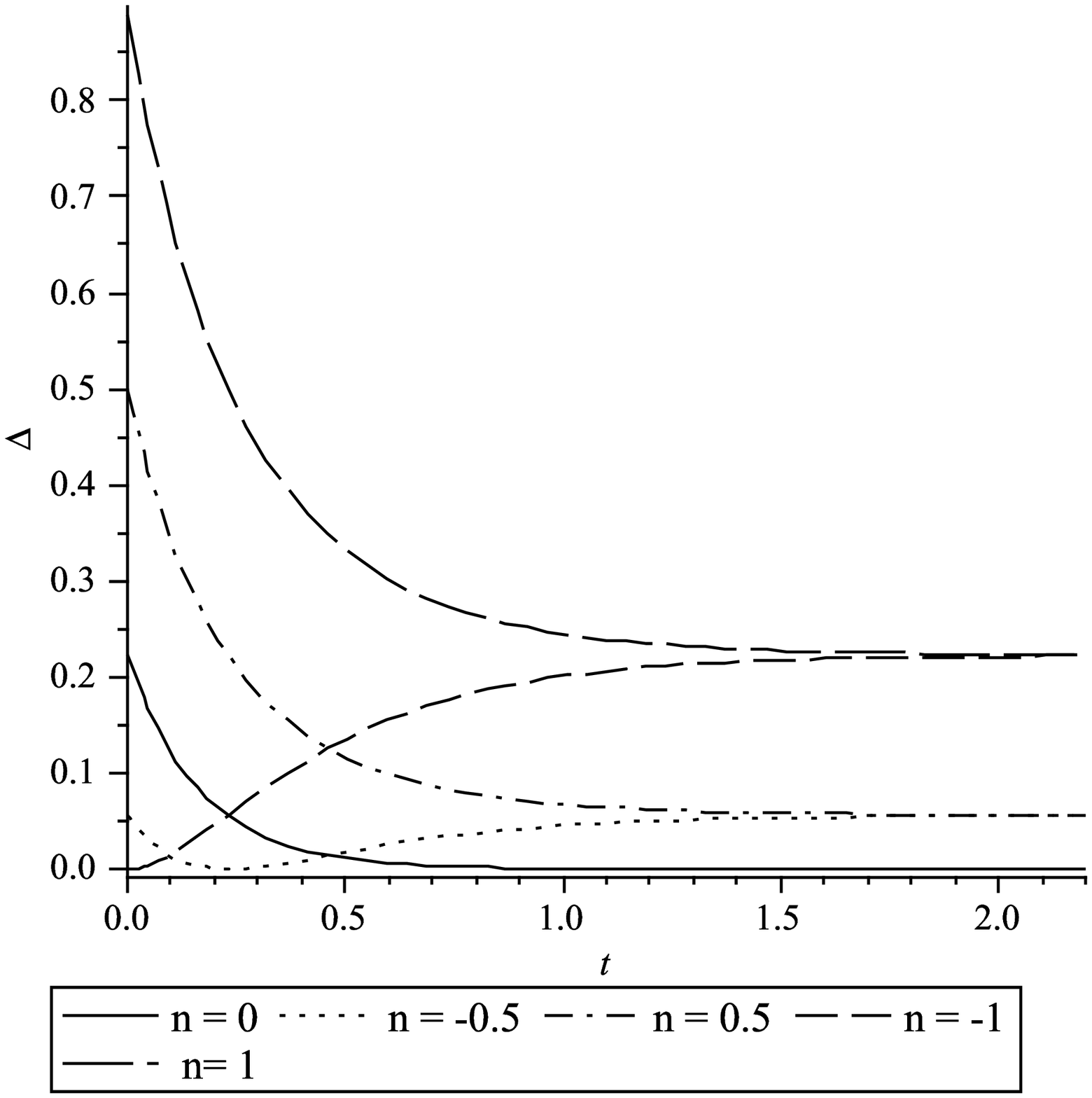}
\caption{$\Delta$ versus cosmic time $t$ for different values of $n$ in model $m=0$. $\lambda$ and $k$ have been chosen as $1$.}
\label{fig:Deltam0}
\end{minipage}
\hspace{0.01\linewidth}
\begin{minipage}[b]{0.49\linewidth}
\centering
\includegraphics[width=1\textwidth]{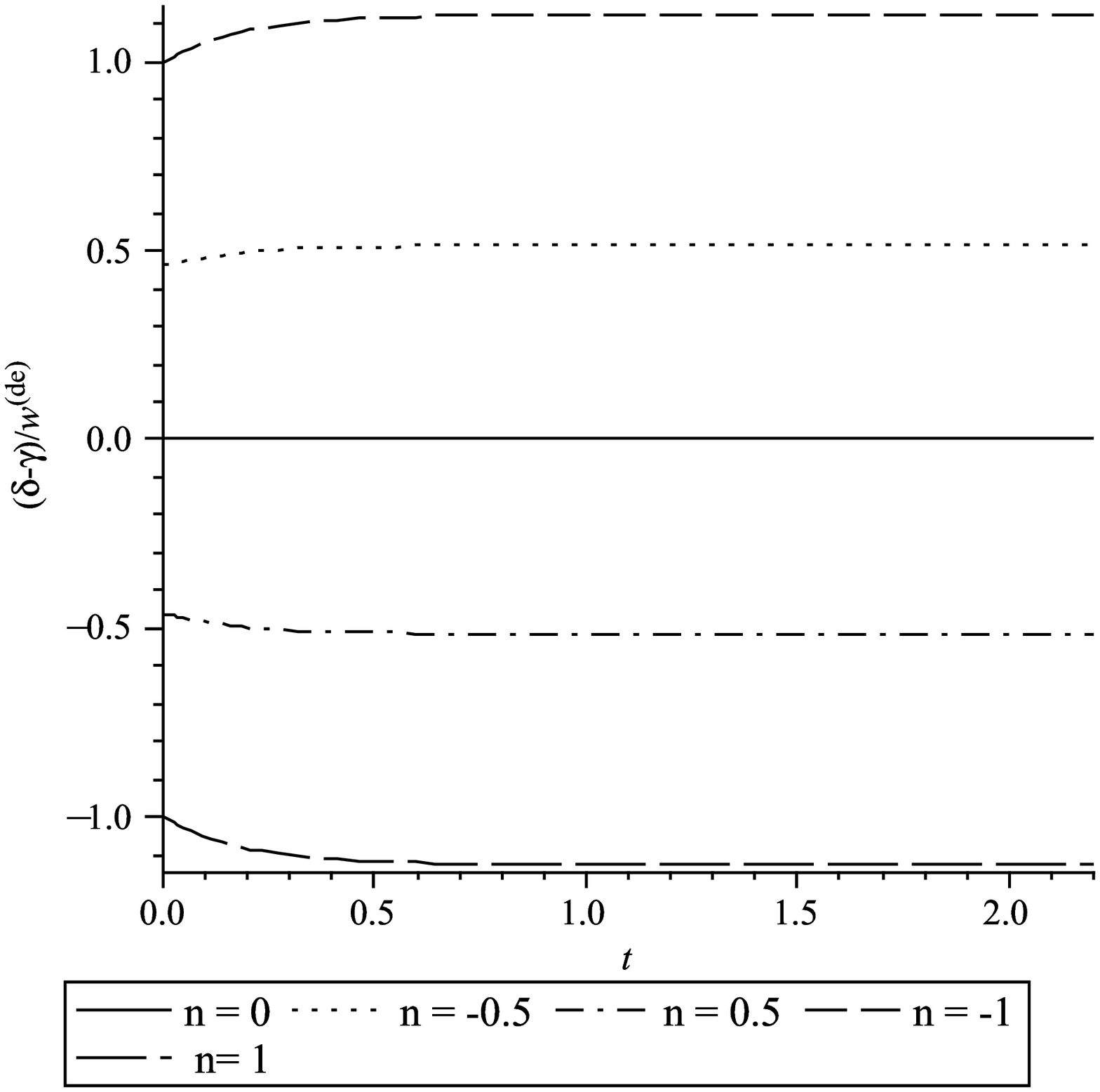}
\caption{The anisotropy of the DE versus cosmic time $t$ for different values of $n$ in model $m=0$. $\lambda$ and $k$ have been chosen as $1$.}
\label{fig:Adem0}
\end{minipage}
\end{figure}

The energy density of the perfect fluid $\rho^{\mathrm{(m)}}$ decreases exponentialy and converges to zero since $w^{\mathrm{(m)}}\geq0$ by definition, while $\rho^{\mathrm{(de)}}$ changes slightly at early times and converges to a non-zero value as $t$ increases. Thus, the ratio of $\rho^{\mathrm{(de)}}/(\rho^{\mathrm{(m)}}+\rho^{\mathrm{(de)}})$ converges to $1$ as $t$ increases, that is the DE dominates the perfect fluid in the inflationary era as expected. Since DE dominates the perfect fluid in the inflationary era as mentioned above, we may neglect the perfect fluid while examining the properties of the DE in this model. $\rho^{\mathrm{(de)}}$ is always positive provided $\vert n+\lambda/k\vert < 3$ and $\vert n \vert <3$, begins with a finite value at $t=0$, exhibits different behaviours depending on the choice of the parameters at the earlier times and tends to $3k^{2}(1-n^{2}/9)$ for large values of $t$; see Fig \ref{fig:rhom0}. It can be observed that the bigger the anisotropy of the expansion the lower the $\rho^{\mathrm{(de)}}$ in any given instant. $\lambda$ loses its effect exponentially as $t$ increases, thus we may say that the higher the $\vert n \vert$ the lower the $\rho^{\mathrm{(de)}}$ in any given instant for relatively big $t$ values. The EoS parameter of the DE $w^{\mathrm{(de)}}$ may begin in phantom ($w<-1$) or quintessence ($w>-1$) region and tends to $-1$ (cosmological constant, $w=-1$) by exhibiting various patterns as $t$ increases; see Fig \ref{fig:wm0}. One can observe that $\rho^{\mathrm{(de)}}$ increases when $w^{\mathrm{(de)}}<-1$, decreases when $w^{\mathrm{(de)}}>-1$ and is constant when $w^{\mathrm{(de)}}=-1$, as would be expected. 

The anisotropy of the expansion decreases monotonically as $t$ increases when $n=0$. However, it exhibits nontrivial behaviour at the early times of the universe and converges to a non-zero value for the late times when $n\neq 0$; see Fig \ref{fig:Deltam0}. The deviation parameters $\delta$ and $\gamma$ are finite at $t=0$, and converge to $2n/(n+3)$ and $n(2n+3)/(n^{2}-9)$ respectively as $t\rightarrow \infty$. The anisotropy of the DE which has been defined as $(\delta-\gamma)/w^{\mathrm{(de)}}$ tends to $9n/(n^{2}-9)$ as $t$ increases; see Fig \ref{fig:Adem0}.
\subsection{Model for $m\neq0\quad(q\neq-1)$}
The solutions in this subsection are valid for all possible values of $m$ except for $m=3$ and $m=0$, thus the solutions for $m=3$ are given in the following subsection.

From (24) one can see that the initial time of the universe is $t_{*}=-c_{2}/mk$ for $m\neq0$. For brevity of the equations, we may redefine the cosmic time as 
\begin{equation}
t'=mkt+c_{2},
\end{equation}
and by doing that the initial time of the universe has also been set to ${t'}=0$. Thus we may rewrite the metric as
\begin{equation}
ds^{2}=(mk)^{-2}dt'^{2}-A(t')^{2}dx^{2}-B(t')^{2}(dy^{2}+dz^{2}).
\end{equation}
Using (31) we may obtain the ratios of the scale factors $A(t)/B(t)$, and on manipulating the result by using (24) we get the following exact expressions for the scale fators:
\begin{equation}
A(t')=\kappa^{2/3}{t'}^{\frac{1}{m}-\frac{2n}{m(m-3)}}e^{\frac{2}{3}\frac{\lambda}{k(m-3)}{t'}^{1-\frac{3}{m}}},
\end{equation}
\begin{equation}
B(t')=\kappa^{-1/3}{t'}^{\frac{1}{m}+\frac{n}{m(m-3)}}e^{-\frac{1}{3}\frac{\lambda}{k(m-3)}{t'}^{1-\frac{3}{m}}},
\end{equation}
where $\kappa$ is the positive constant of integration. The spatial volume of the universe is found as
\begin{equation}
V={t'}^{\frac{3}{m}}.
\end{equation}
The directional Hubble parameters as defined in (21) are found as
\begin{equation}
H_{x}=k{t'}^{-1}+\frac{2}{3}{\lambda}{t'}^{-\frac{3}{m}}-\frac{2nk}{(m-3)}{t'}^{-1},
\end{equation}
\begin{equation}
H_{y,z}=k{t'}^{-1}-\frac{1}{3}{\lambda}{t'}^{-\frac{3}{m}}+\frac{nk}{(m-3)}{t'}^{-1}.
\end{equation}
Using (26), (52) and (53) in (38) we get
\begin{equation}
\Delta(t')=\frac{2}{9}\left(\frac{\lambda}{k}{t'}^{1-\frac{3}{m}}-\frac{3n}{m-3}\right)^{2},
\end{equation}
for the anisotropy parameter of the expansion. The expansion and shear scalars are, respectively, found as
\begin{equation}
\theta =3k{t'}^{-1}=3H,
\end{equation}
\begin{equation}
\sigma^{2}=\frac{1}{3}k^{2}{t'}^{-2}\left(\frac{\lambda}{k}{t'}^{1-\frac{3}{m}}-\frac{3n}{m-3}\right)^{2}.
\end{equation}
Using the scale factors (49-50) in (11), the energy density of the perfect fluid is found as
\begin{equation}
\rho^{\mathrm{(m)}}(t') ={\rho^{\mathrm{(m)}}_{0}}{t'}^{-\frac{3}{m}(1+w^{\mathrm{(m)}})}.
\end{equation}
The energy density of the DE can be found from (5) by using the scale factors (49-50) and the energy density of the perfet fluid (57), and may be written in terms of $\Delta(t)$ as
\begin{equation}
\rho^{\mathrm{(de)}}(t')=3k^{2}\left(1-\frac{1}{2}\Delta(t')\right){t'}^{-2}-\rho^{\mathrm{(m)}}(t').
\end{equation}
Using (49-50) and (58) in (14) we get
\begin{equation}
w^{\mathrm{(de)}}(t')=\frac{\frac{2}{3}\frac{m}{m-3}\lambda nk{t'}^{-1-\frac{3}{m}}-\frac{1}{3}{\lambda}^{2}{t'}^{-\frac{6}{m}}+(2m-3)k^{2}{t'}^{-2}\left(1-\frac{n^{2}}{(m-3)^{2}}\right)-w^{\mathrm{(m)}}\rho^{\mathrm{(m)}}(t')}{3k^{2}\left(1-\frac{1}{2}\Delta(t')\right){t'}^{-2}-\rho^{\mathrm{(m)}}(t')},
\end{equation}
for the deviation-free EoS parameter of the DE. And finally using equations (49-50) and (58) in (15) and (16) we may get deviation parameters as following,
\begin{equation}
\delta(t')=\frac{nk{t'}^{-2}\left(2k-\frac{2}{3}{\lambda}{t'}^{1-\frac{3}{m}}+\frac{2nk}{(m-3)}\right)}{3k^{2}\left(1-\frac{1}{2}\Delta(t')\right){t'}^{-2}-\rho^{\mathrm{(m)}}(t')},
\end{equation}
\begin{equation}
\gamma(t')=\frac{nk{t'}^{-2}\left(-k-\frac{2}{3}{\lambda}{t'}^{1-\frac{3}{m}}+\frac{2nk}{(m-3)}\right)}{3k^{2}\left(1-\frac{1}{2}\Delta(t')\right){t'}^{-2}-\rho^{\mathrm{(m)}}(t')}.
\end{equation}
\subsection{Physical behaviour of the model for $m\neq0\quad(q\neq-1)$}
The universe accelerates for $0<m<1$, decelerates for $m>1$ and expands with constant velocity for $m=1$. This model may represent the radiation dominated era for $m=2$ and $w^{\mathrm{(m)}}=1/3$, and the matter dominated era for $m=3/2$ and $w^{\mathrm{(m)}}=0$.

The mean Hubble parameter $H$, expansion scalar $\theta$ and shear scalar $\sigma^{2}$ are infinitly large at $t'=0$, and null at $t'=\infty$.

If $m>3$, the space time exhibits a cigar type singularity at $t'=\infty$ for $\lambda>0$ and a pancake type singularity for $\lambda<0$. If $m<3$, all the axes expand to infinitly large values provided $(m-3)/2<n<3-m$ as $t'\rightarrow\infty$. When $n=3-m$, $A(t')$ takes infinitly large values as $t'\rightarrow\infty$, while $B(t')$ converges to $\kappa^{-1/3}$. When $n=(m-3)/2$, $B(t')$ takes infinitly large values as $t'\rightarrow\infty$, while $A(t')$ converges to $\kappa^{2/3}$. When $n>3-m$ space time exhibits a cigar type singularity at $t'=\infty$. When $n<(m-3)/2$ space time exhibits a pancake type singularity at $t'=\infty$.

$\Delta$ diverges as $t'\rightarrow 0$, converges to a constant as $t'\rightarrow \infty$ for $m<3$ and vice versa for $m>3$. One can observe that the anisotropy of the expansion lowers the value of $\rho^{\mathrm{(de)}}$. Thus, $\Delta$ must be considered while examining the dynamics of the energy density of the DE, particularly since $\Delta$ diverges in the above mentioned limits. In the model for $m<3$, $\rho^{\mathrm{(de)}}$ will eventually obtain negative values as $t'$ goes to zero due to the divergence of $\Delta$. Hence, this model is not appropriate for representing the relatively earlier times of the universe. On the other hand, this model can set to represent relatively later times of the universe since we can always set $\rho^{\mathrm{(de)}}$ to be $\rho^{\mathrm{(de)}}\geq0$ as $t'\rightarrow \infty$ by choosing the parameters in suitable intervals. Similarly, in the model for $m>3$, $\rho^{\mathrm{(de)}}$ will eventually obtain negative values as $t'\rightarrow \infty$. Hence, this model is not appropriate for representing the late times of the universe, but for the earlier times of the universe by choosing the parameters in suitable intervals.

$w^{\mathrm{(de)}}$ and $(\delta-\gamma)/w^{\mathrm{(de)}}$, say the anisotropy of the DE, exhibits various dynamics according to the choice of the parameters. It is worth to mention here once more that the anisotropy of the DE doesn't always act so as to increase the anisotropy of the expansion, when the signs of $\lambda$ and $n$ are opposite the overall effect is to lower the expansion anisotropy.

\begin{figure}[ht]
\begin{minipage}[b]{0.49\linewidth}
\centering
\includegraphics[width=1\textwidth]{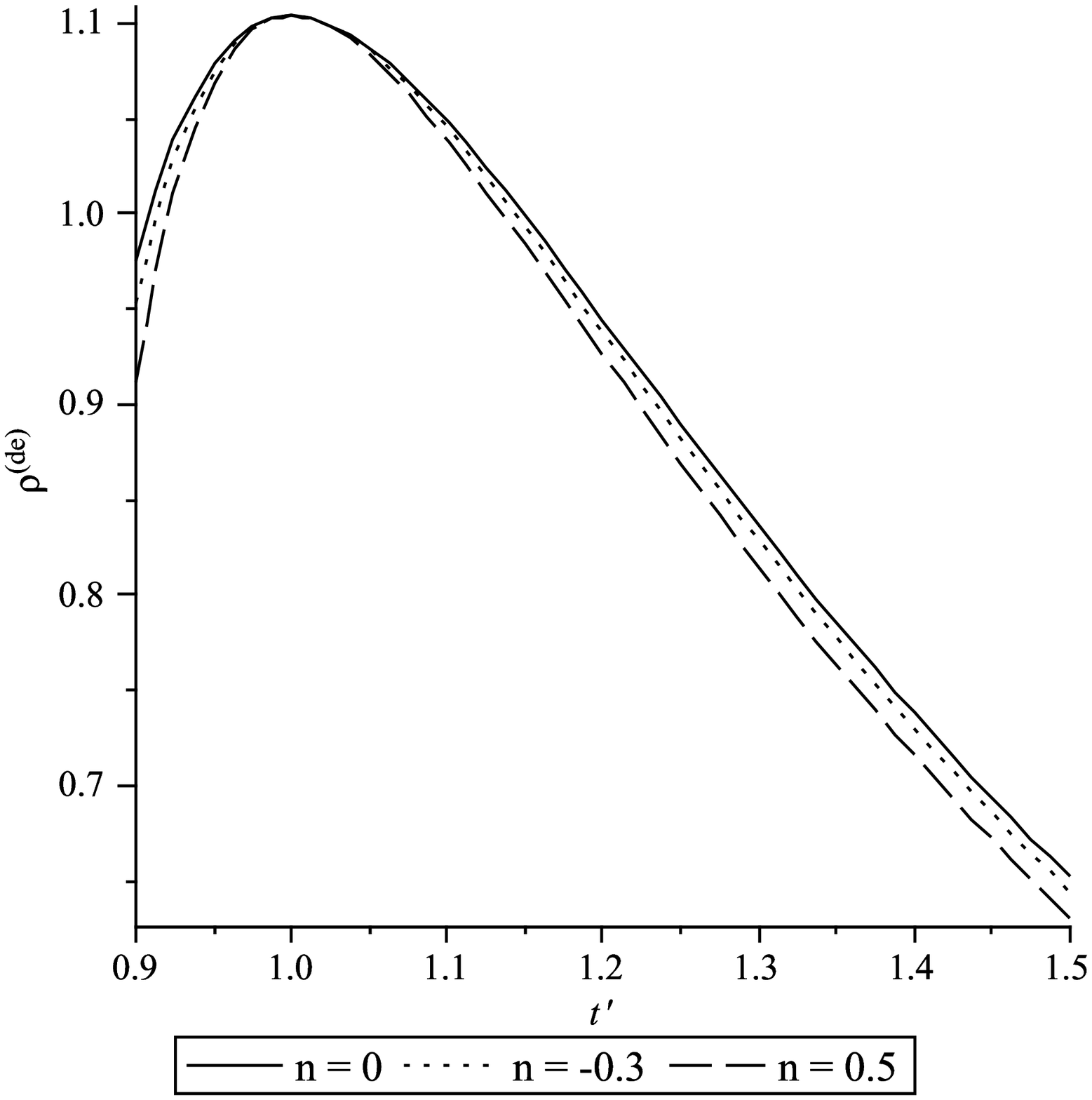}
\caption{$\rho^{\mathrm{(de)}}$ versus cosmic time $t'$ in the vicinity of $t'=1$ in the model $m=0.405$ for different values of $n$. $\Delta$ has been set to null at $t'=1$ by choosing $\lambda=-0.8208092487n$.}
\label{fig:rhodem}
\end{minipage}
\hspace{0.01\linewidth}
\begin{minipage}[b]{0.49\linewidth}
\centering
\includegraphics[width=1\textwidth]{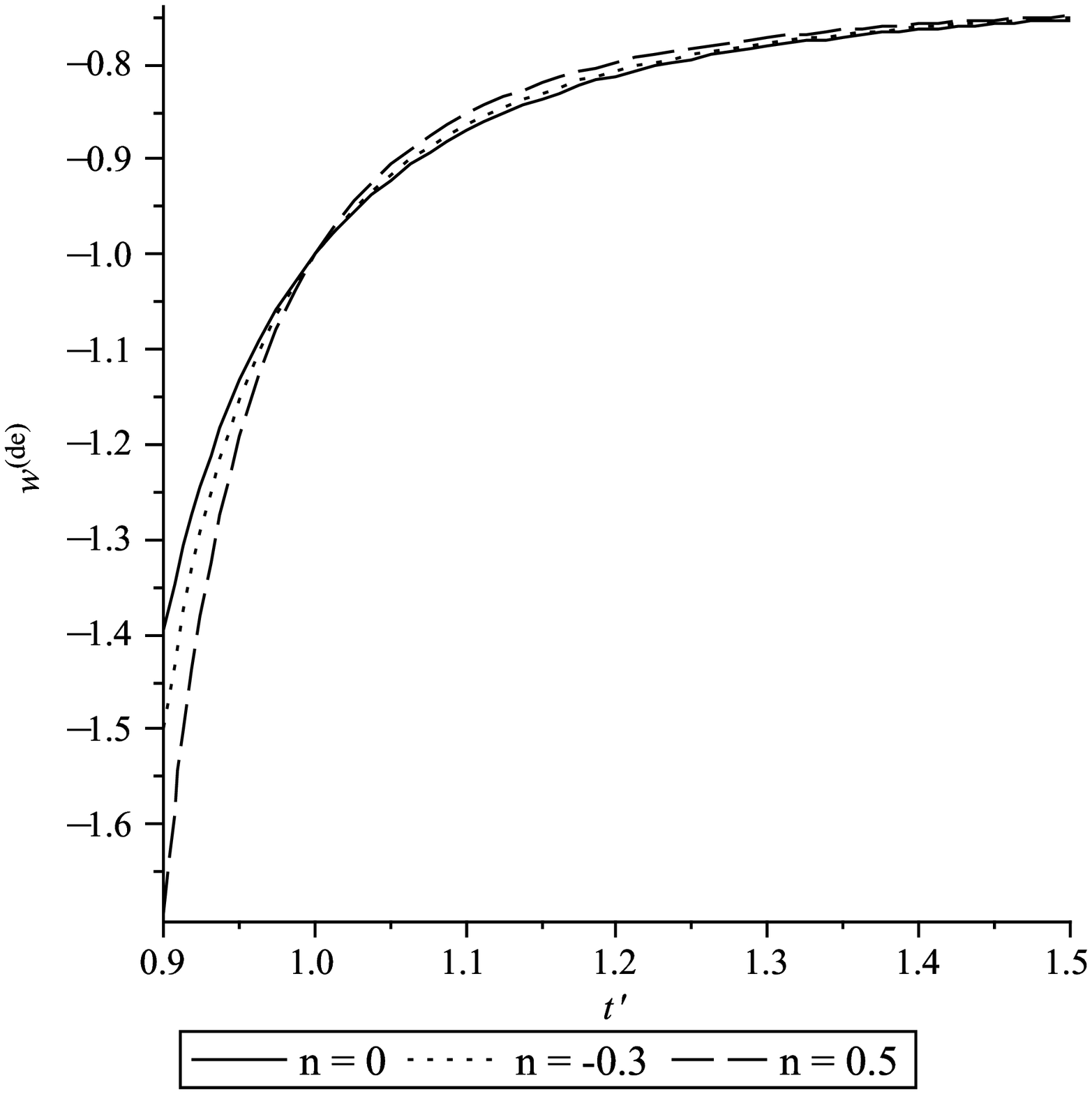}
\caption{The plot of $w^{\mathrm{(de)}}$ versus cosmic time $t'$ in the vicinity of $t'=1$ in the model $m=0.405$ for different values of $n$. $\Delta$ has been set to null at $t'=1$ by choosing $\lambda=-0.8208092487n$. $w^{\mathrm{(de)}}(1)=-1$}
\label{fig:wm}
\end{minipage}
\end{figure}
\begin{figure}[ht]
\begin{minipage}[b]{0.49\linewidth}
\centering
\includegraphics[width=1\textwidth]{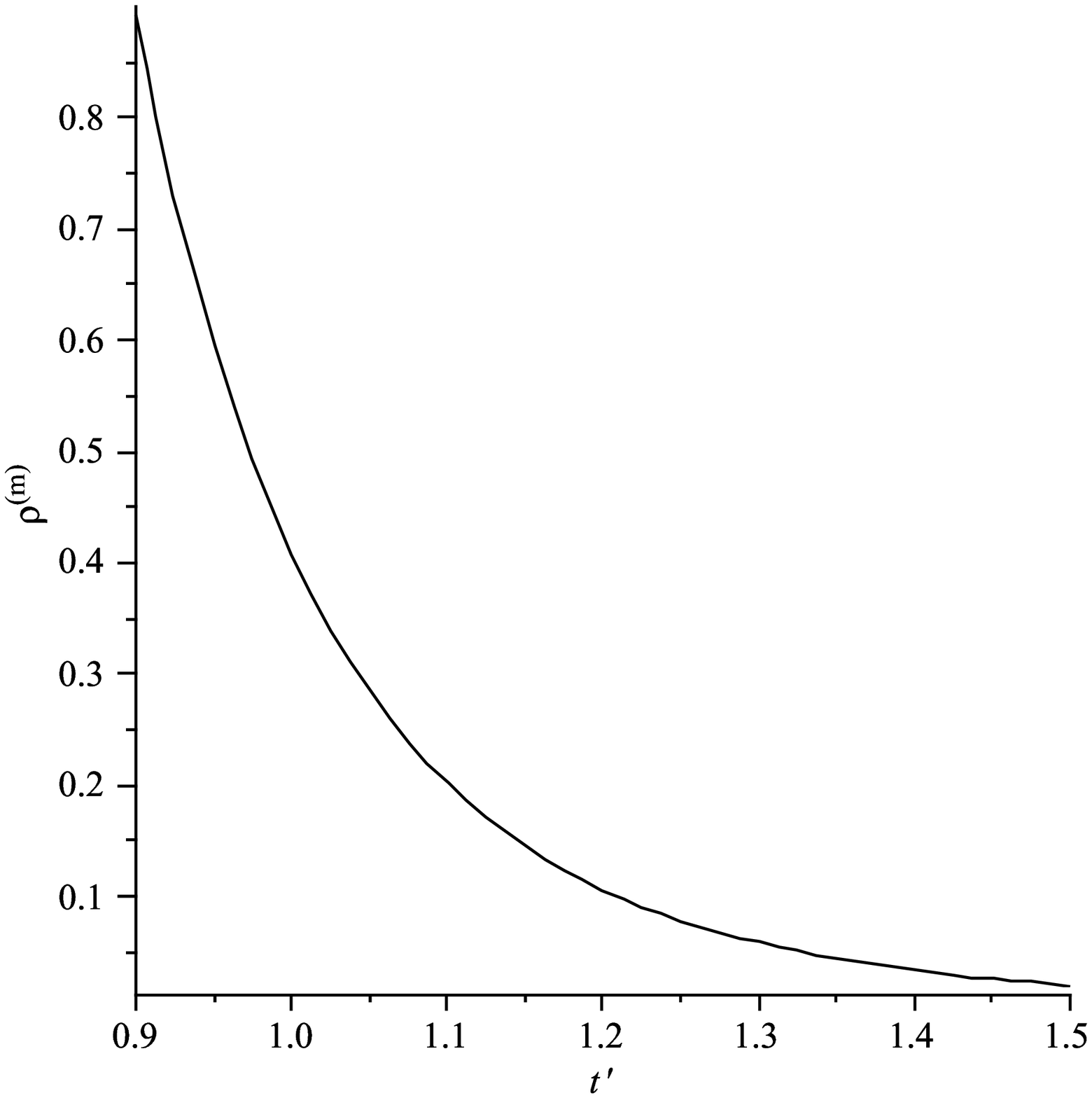}
\caption{The plot of $\rho^{\mathrm{(m)}}$ with $w^{\mathrm{(m)}}=0$, versus cosmic time $t'$ in the vicinity of $t'=1$ in the model $m=0.405$. The behaviour of the $\rho^{\mathrm{(m)}}$ is indepedent of $n$.}
\label{fig:rhomm}
\end{minipage}
\hspace{0.01\linewidth}
\begin{minipage}[b]{0.49\linewidth}
\centering
\includegraphics[width=1\textwidth]{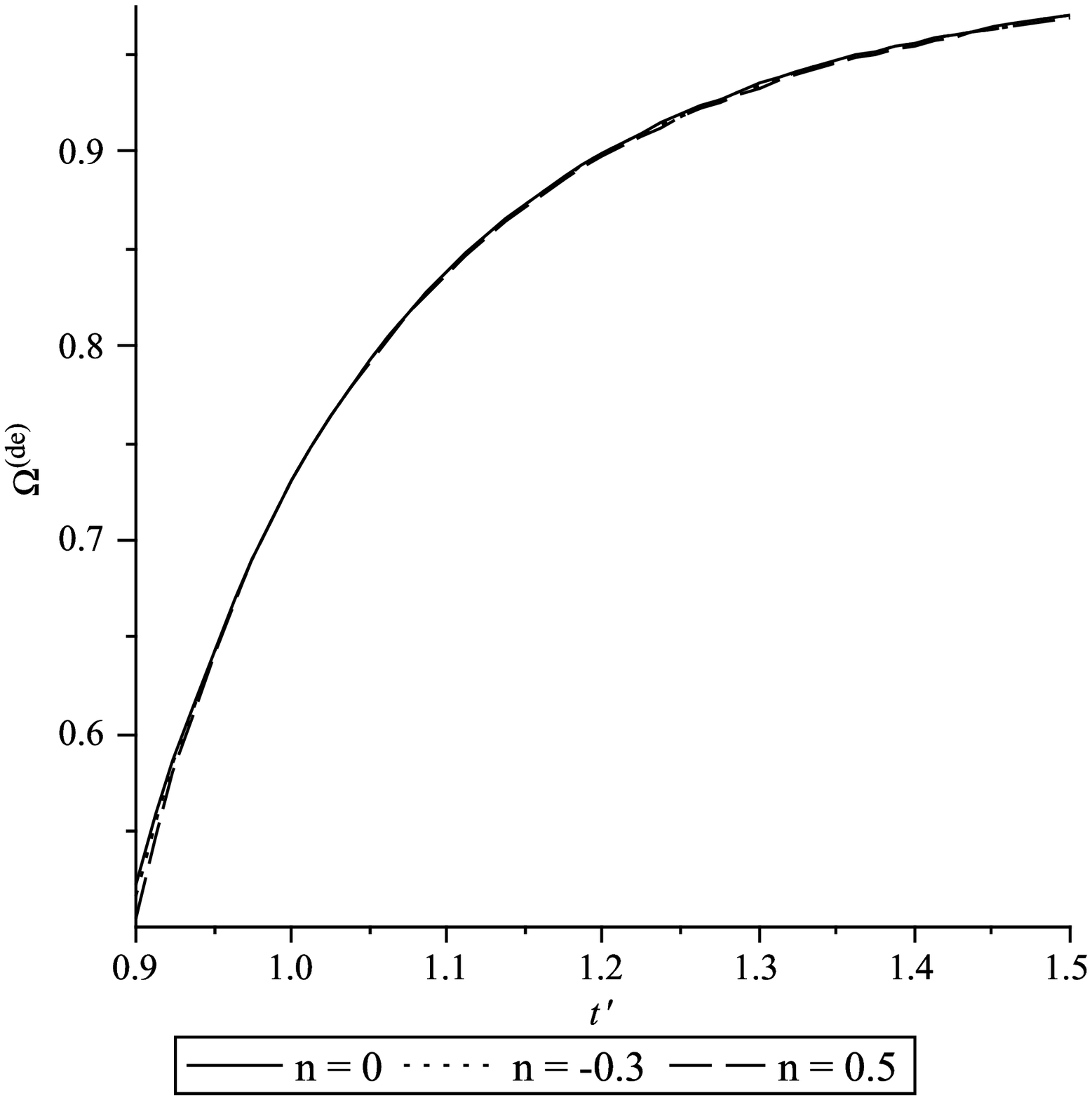}
\caption{The plot of $\Omega^{\mathrm{(de)}}$ versus cosmic time $t'$ in the vicinity of $t'=1$ in the model $m=0.405$ for different values of $n$. $\Delta$ has been set to null at $t'=1$ by choosing $\lambda=-0.8208092487n$. $\Omega^{\mathrm{(de)}}(1)=0.73$}
\label{fig:Omegam}
\end{minipage}
\end{figure}
\begin{figure}[ht]
\begin{minipage}[b]{0.49\linewidth}
\centering
\includegraphics[width=1\textwidth]{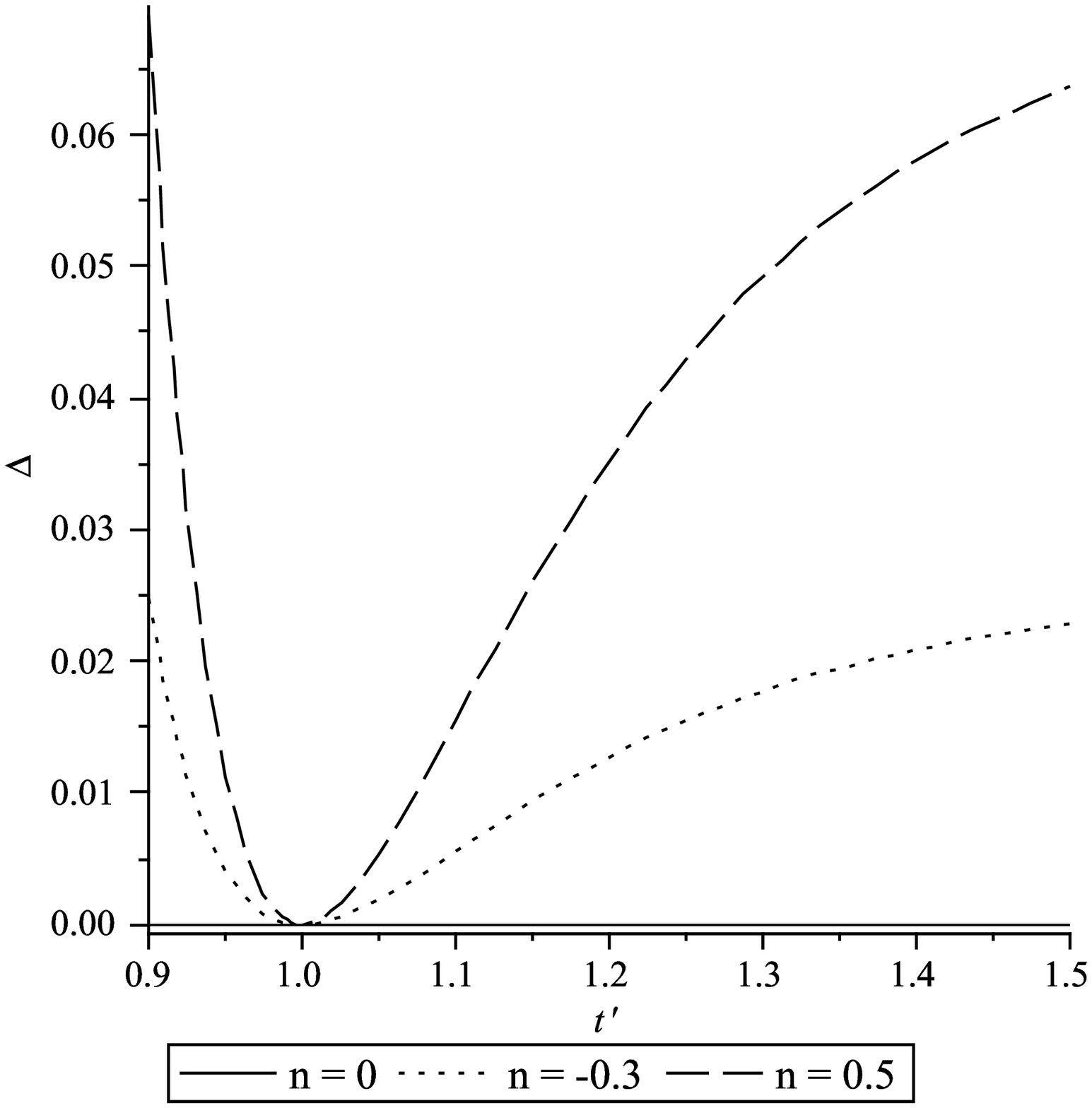}
\caption{The plot of $\Delta$ versus cosmic time $t'$ in the vicinity of $t'=1$ in the model $m=0.405$ for different values of $n$. $\Delta$ has been set to null at $t'=1$ by choosing $\lambda=-0.8208092487n$.}
\label{fig:Deltam}
\end{minipage}
\hspace{0.01\linewidth}
\begin{minipage}[b]{0.49\linewidth}
\centering
\includegraphics[width=1\textwidth]{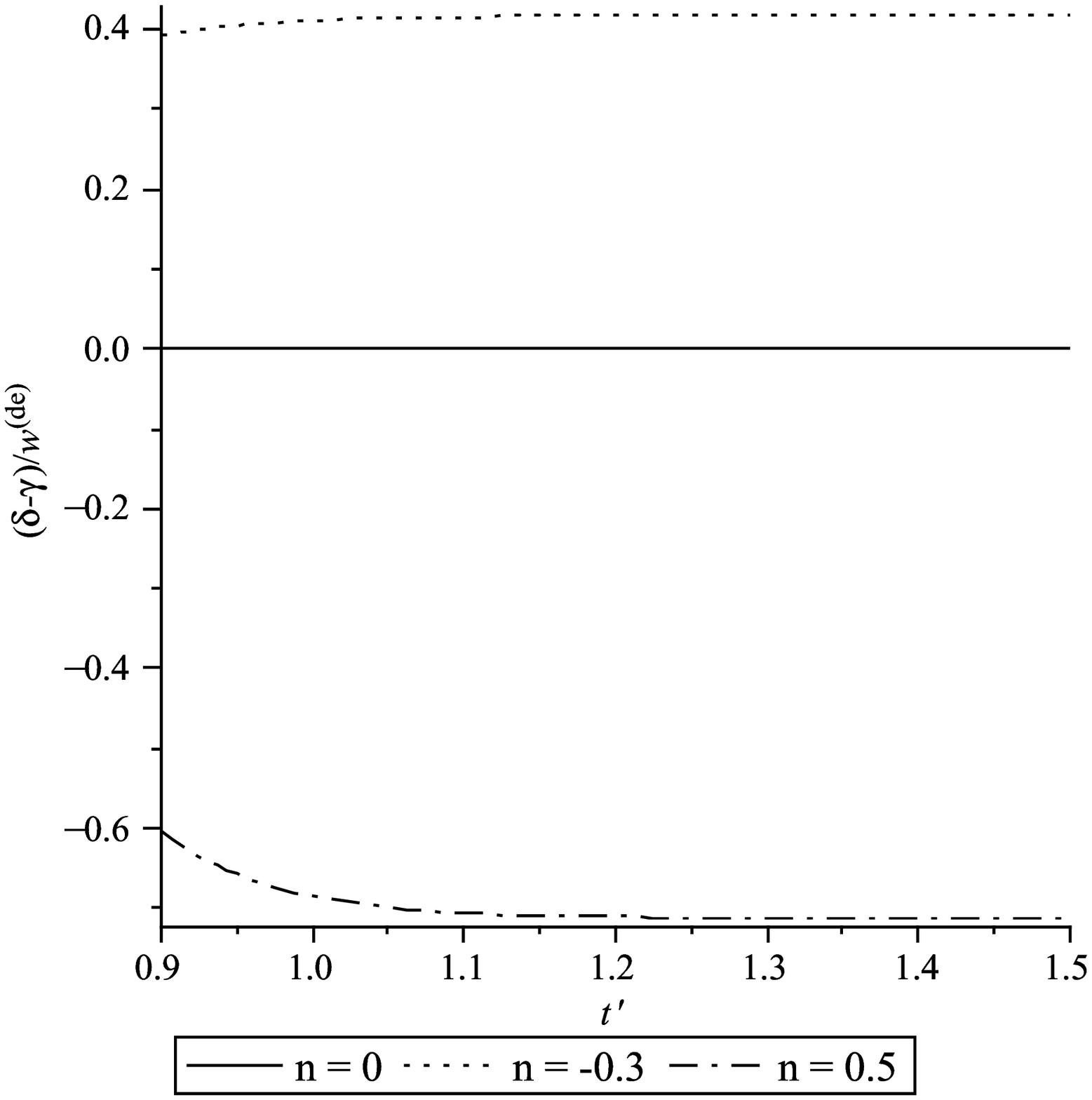}
\caption{Plot of the anisotropy of the DE versus cosmic time $t'$ in the vicinity of $t'=1$ in the model $m=0.405$ for different values of $n$. $\Delta$ has been set to null at $t'=1$ by choosing $\lambda=-0.8208092487n$.}
\label{fig:adem}
\end{minipage}
\end{figure}
We may try to set $t'=1$ to the present universe by using the recent cosmological data which favor $H_{0}\approx0.71$ (dimensionless Hubble parameter), $\Omega^{\mathrm{(de)}}_{0}\approx0.73$ (density parameter of the dark energy), $\Omega^{\mathrm{(m)}}_{0}\equiv1-\Omega^{\mathrm{(de)}}_{0}\approx0.27$, $w^{\mathrm{(de)}}_{0}\approx-1$ \cite{{Hinshaw08}}, $w^{\mathrm{(m)}}=0$ and a space-time that can be represented by flat RW metric which implies $\Delta_{0}=0$ for the present universe, where $_{0}$ subscripts represent present ($t'=1$) values of the parameters. With above parameters we may obtain present critical energy density of the universe as $\rho^{\mathrm{(c)}}_{0}=1.5123$, the deceleration parameter as $q_{0}=-0.595$ which implies $m=0.405$ in our model, $\rho^{\mathrm{(m)}}_{0}=0.408321$, $\rho^{\mathrm{(de)}}_{0}=1.103979$, $w^{\mathrm{(de)}}_{0}=-1$ and $\Delta_{0}=0$ by choosing $\lambda=-0.8208092487n$.

In this model with the above values of the parameters, which may correspond to the present universe in the vicinity of $t'=1$, $\rho^{\mathrm{(m)}}$ is roughly proportional with $t'^{-8}$, while $\rho^{\mathrm{(de)}}$ is roughly proportional with $t^{-2}$ and gets its highest value at $t'=1$ where $\Delta=0$. That is, the DE eventually dominates the perfect fluid. Some graphs may be plotted with above values, since we can think that $q$ is changing only slightly for the present universe, for $n=0$, $n=-0.3$ and $n=0.5$. Plots has been extended towards to the later times more than towards to the earlier times since, as mentioned above, models with $m<3$ are not appropriate for representing relatively earlier times of the universe. $\rho^{\mathrm{(de)}}$ increases when $w^{\mathrm{(de)}}$ is in the phantom region, begins to decrease when $w^{\mathrm{(de)}}$ passes into the quintessence region and converges to zero as $t'\rightarrow \infty$; see Fig \ref{fig:rhodem}. $w^{\mathrm{(de)}}$ begins in the phantom region, increases and becomes $-1$ at $t'=1$ then passes into the quintessence region and tends to a constant which is in the quintessence region; see Fig \ref{fig:wm}. $\rho^{\mathrm{(m)}}$ decreases as $t'$ increases independent of $n$; see Fig \ref{fig:rhomm}. $\Omega^{\mathrm{(de)}}$ increases as $t'$ increases, becomes $0.73$ at $t'=1$ and converges to $1$ as $t'$ keeps on increasing; see Fig \ref{fig:Omegam}. $\Delta$ decreases as $t'$ increases and becomes null at $t'=1$ then tends to a constant which depends on $n$, see Fig \ref{fig:Deltam}. The bigger the $\Delta$ the lower the $\rho^{\mathrm{(de)}}$, consequently lower the $\Omega^{\mathrm{(de)}}$ in any given instant. The anisotropy of the DE, $(\delta-\gamma)/w^{\mathrm{(de)}}$, is null for $n=0$, changes slightly and converges to a non-zero constant for $n\neq0$; see Fig \ref{fig:adem}.
\subsection{Model for $m=3\quad(q=2)$}
From (24) we can see that the initial time of the universe is $t_{*}=-c_{2}/3k$ for $m=3$. For brevity of the equations, we may redefine the cosmic time as 
\begin{equation}
t'=3kt+c_{2},
\end{equation}
and by doing that the initial time of the universe has also been set as ${t'}=0$. Thus we may rewrite the metric as
\begin{equation}
ds^{2}=(3k)^{-2}dt'^{2}-A(t')^{2}dx^{2}-B(t')^{2}(dy^{2}+dz^{2}).
\end{equation}
Using (32) we may get the ratios of the scale factors $A(t)/B(t)$ for $m=3$, and manipulating the result by using (24) we get the following exact expressions for the scale fators;
\begin{equation}
A(t')=\kappa^{\frac{2}{3}}{t'}^{\frac{1}{3}+\frac{2}{9}\frac{\lambda}{k}}e^{\frac{n}{9}{\ln(t')}^{2}},
\end{equation}
\begin{equation}
B(t')=\kappa^{-\frac{1}{3}}{t'}^{\frac{1}{3}-\frac{1}{9}\frac{\lambda}{k}}e^{-\frac{n}{18}{\ln(t')}^{2}},
\end{equation}
where $\kappa$ is the positive constant of integration. The spatial volume of the universe is found as
\begin{equation}
V={t'}.
\end{equation}
The directional Hubble parameters as defined in (21) are found as
\begin{equation}
H_{x}=k{t'}^{-1}+\frac{2}{3}\left(\lambda +nk\ln({t'})\right){t'}^{-1},
\end{equation}
\begin{equation}
H_{y,z}=k{t'}^{-1}-\frac{1}{3}\left(\lambda +nk\ln({t'})\right){t'}^{-1}.
\end{equation}
Using (26) and the scale factors in (38) we get
\begin{equation}
\Delta=\frac{2}{9}k^{-2}\left(nk\ln(t')+\lambda\right)^2
\end{equation}
for the anisotropy parameter of expansion. The expansion and shear scalars are, respectively, found as
\begin{equation}
\theta =3k{t'}^{-1}=3H,
\end{equation}
\begin{equation}
\sigma^{2}=\frac{1}{3}\left(nk\ln(t')+\lambda\right)^2{t'}^{-2}.
\end{equation}
Using the scale factors in (11), the energy density of the perfect fluid is found as
\begin{equation}
\rho^{\mathrm{(m)}}(t') ={\rho^{\mathrm{(m)}}_{0}}{t'}^{-(1+w^{\mathrm{(m)}})}.
\end{equation}
The energy density of the DE can be found from (5) by using the scale factors and the energy density of the perfet fluid (72),
\begin{equation}
\rho^{\mathrm{(de)}}(t')=3k^{2}\left(1-\frac{1}{2}\Delta(t')\right){t'}^{-2}-\rho^{\mathrm{(m)}}(t'),
\end{equation}
Using (64-65) and (73) in (14) we get
\begin{equation}
w^{\mathrm{(de)}}(t')=\frac{3k^{2}{t'}^{-2}-\frac{1}{3}(nk\ln({t'})+\lambda)^{2}{t'}^{-2}+\frac{2}{3}nk(nk\ln{(t')}+\lambda){t'}^{-2}-w^{\mathrm{(m)}}\rho^{\mathrm{(m)}}(t)}{3k^{2}\left(1-\frac{1}{2}\Delta(t')\right){t'}^{-2}-\rho^{\mathrm{(m)}}(t')}
\end{equation}
for the deviation-free EoS parameter of the DE. And finally using equations (64-65) and (73) in (15) and (16) we may get deviation parameters as following,
\begin{equation}
\delta(t')=\frac{nk\left(2k-\frac{2}{3}nk\ln({t'})-\frac{2}{3}\lambda\right){t'}^{-2}}{3k^{2}\left(1-\frac{1}{2}\Delta(t')\right){t'}^{-2}-\rho^{\mathrm{(m)}}(t')},
\end{equation}
\begin{equation}
\gamma(t')=\frac{nk\left(-k-\frac{2}{3}nk\ln({t'})-\frac{2}{3}\lambda\right){t'}^{-2}}{3k^{2}\left(1-\frac{1}{2}\Delta(t')\right){t'}^{-2}-\rho^{\mathrm{(m)}}(t')}.
\end{equation}
\subsection{Physical behaviour of the model for $m=3\quad(q=2)$}
Nothing physically special has been observed in this model for further investigation. But we may mention that this model may only be valid for intermediate epochs of the universe. Because $\Delta$ diverges at $t'=0$ and $t'=\infty$ in this model, thus $\rho^{\mathrm{(de)}}$ will eventually get negative values as $t'$ goes to extrem values, since the anisotropy of the expansion contribute the energy density of the DE negatively.
 
%as required. Don't forget to give each section
%and subsection a unique label (see Sect.~\ref{sec:1}).
%\paragraph{Paragraph headings} Use paragraph headings as needed.
\section{Conclusion}
Locally rotationally symmetric Bianchi-I cosmological models with dynamically anisotropic dark energy (DE) and perfect fluid have been constructed in General Relativity. We assume that the dark energy (DE) is minimally interacting, has dynamical energy density and anisotropic equation of state parameter (EoS). The conservation of the energy-momentum tensor of the DE has been assumed to consist of two separately additive conserved parts. A special law has been assumed for the deviation from isotropic EoS, which is consistent with the assumption on the conservation of the energy-momentum tensor of the DE. Exact solutions of Einstein's field equations have been obtained by assuming a special law of variation for the mean Hubble parameter, which yields a constant value of the deceleration parameter and is not inconsistent with observations. Some basic geometrical and kinematical features of the models and the dynamics of the anisotropic DE in these models have been examined.

A more general EoS parameter has been introduced for the DE, and the isotropic DE can be recovered by choosing $n$ to be null, where $n$ parametrizes the amplitude of the deviation from isotropic EoS parameter of the DE. In all the models, while the anisotropy of the DE contributes to the expansion of one of the scale factors, it opposes to the expansion of the other. Two parameters that parametrize the difference between directional Hubble parameters,  $\lambda$ and $n$, give rise to non-trivial, dynamically anisotropic, expansion histories. This allows for the possibility to fine tune the isotropization of the Bianchi metric during an accelerating epoch of the universe in order to generate arbitrary ellipsoidality by choosing a suitable value of $n$. Such a result provides also the possibility to fine tune the CMB anisotropy \cite{{Campanelli06},{Campanelli07}}. The anisotropy of the expansion can mildly or totally isotropize in relatively earlier times of the universe. Nevertheless, it converges to a nonzero constant value for the later times of the universe in all models. It is also observed that the anisotropy of the DE energy doesn't always act so as to increase the anisotropy of the expansion, when signs of $\lambda$ and $n$ are opposite the overall effect is to lower the expansion anisotropy in relatively earlier times of the universe.

One interesting observation is that the higher the anisotropy of the expansion the lower the energy density of the DE in each given instant of the cosmic time. $w^{\mathrm{(de)}}$ is also dynamical and exhibits non-trivial behaviour in our model, but it eventually converges to a constant in the quintessence region for $m \neq 0$ and converges to $-1$ for $m=0$.

Such cosmological models are of interest because they give rise to an ellipsoidality of the universe in spite of the inflation \cite{{Rodrigues},{Koivisto08a},{Koivisto08b}}, which is one of the promising proposals to the solution of the quadrupole problem \cite{{Eriksen},{Copi},{Campanelli06},{Campanelli07}} and can also be checked by the direction dependency of the redshift-luminosity relation of the SNIa observations \cite{{Koivisto08a},{Koivisto08b}}.
%\begin{figure}
% \includegraphics{example.eps}
% \caption{Please write your figure caption here}
%\end{figure}
%\begin{figure*}
% \includegraphics[width=0.75\textwidth]{deneme.eps}
% \caption{Please write your figure caption here}
%\end{figure*}
%\begin{table}
%\begin{tabular}{lll}
%\hline\noalign{\smallskip}
%\end{tabular}
%\end{table}
%\begin{acknowledgements}
%\end{acknowledgements}

\end{document}